\documentclass[article]{revtex4}
\usepackage{pdfpages}
\usepackage{epsf}
\usepackage{subfigure}
\usepackage{amsmath}
\usepackage{eqnarray,amsmath}
\usepackage{epstopdf}
\usepackage{mathrsfs}
\usepackage{natbib}
\usepackage{amssymb,latexsym}
\usepackage{dcolumn}
\usepackage{graphicx}
\usepackage{color}

\makeatletter
\newcommand*{\rom}[1]{\expandafter\@slowromancap\romannumeral #1@}
\makeatother
\def\be{\begin{equation}}
\def\ee{\end{equation}}
\def\ba{\begin{eqnarray}}
\def\ea{\end{eqnarray}}

\graphicspath{{images/}}
\begin{document}
	\title{\large \bf  Quantum decoherence from entanglement during inflation}
	
	\author{Abasalt Rostami}
	\affiliation{Department of Physics, Sharif University of Technology,
		Tehran, Iran }
	\affiliation{ School of Physics, Institute for Research in Fundamental Sciences (IPM), P. O. Box 19395-5531, Tehran, Iran }
	\email{aba-rostami@ipm.ir}
	
	\author{Javad T. Firouzjaee}
	\affiliation{ School of Astronomy, Institute for Research in Fundamental Sciences (IPM), P. O. Box 19395-5531, Tehran, Iran }
	\email{j.taghizadeh.f@ipm.ir}
	\begin{abstract}
		
We study the primary entanglement effect on the decoherence of fields  reduced density matrix which are in interaction with another fields or independent mode functions. We show that the primary entanglement has a significant role in decoherence of the system quantum state. We find that the existence of entanglement could couple  dynamical equations coming from Schr\"{o}dinger equation. We show if one wants to see no effect of the entanglement parameter in decoherence then interaction terms in Hamiltonian can not be independent from each other. Generally, including the primary entanglement destroys the independence of the interaction terms. Our results could be generalized to every scalar quantum field theory with a well defined quantization in a given curved space time.
		
	\end{abstract}
	%
	%
	\maketitle
	\tableofcontents
	
\newpage
	
	\section{Introduction}

Recent Cosmic Microwave Background (CMB)  observations \cite{planck} of temperature fluctuations agree well  with the predictions made by inflation theory which states the universe was in a accelerated phase after the Planck era.  According to the inflationary picture, not only primordial cosmological  fluctuations (CMB temperature) have quantum origin, but they are also created in a quantum state. This quantum fluctuation provides an elegant description for the large scale structure advent of our universe which explain how the density perturbations which seeded structures in cosmos.\\

The basic point is that no noted cosmological data would present the actual  quantum state of primordial fluctuations because all known techniques of observation focus on  a restricted set of properties of those fluctuations.  Consequently, one has to consider not  only the quantum aspects of the cosmological fluctuations, but also the loss of quantum  coherence prompted by the partial description appropriate observation to study nontrivial quantum behavior.\\

Along this way, main work on the quantum to classical transition of inflation \cite{Polarski:1995jg,Albrecht:1992kf,Kiefer:2008ku,Martin:2015qta} has focused largely on the squeezing of the quantum state for each mode. This squeezing, present on super-Hubble scales at the end of inflation, means that the
inflaton is effectively taking on different values at widely separated points in space, which
leads to an inhomogeneity in the temperature after  inflation which means quantum expectation values
of products of quantities in a highly squeezed state bring out to be identical to stochastic
averages calculated from a stochastic distribution of classical quantity configurations.\\

However, this quantum to classical transition does not answer the question of how we make a measurement. As is known, the inflaton field fluctuation is a quantum field which is represented by Bunch-Davies vacuum state which is completely homogeneous and isotropic. The inflaton dynamics preserves the homogeneity and isotropy. Thus, we can not use inflaton state to explain the observed inhomogeneous and anisotropic distribution of the primordial energy density in our universe. As a result, the homogeneous quantum state which is a coherent superposition of all field configurations
collapses to a particular stochastic realization of classical inhomogeneities \cite{Leon:2015eza,Leon:2016mdn}. In this step, it is needed
to have a mechanism of quantum decoherence which necessitates the presence of additional environment degrees of freedom that couple to quantum
perturbations as a measuring device. Decoherence describes the transition from pure
to mixed state which arises whenever the degrees of freedom of interest (quantum fluctuations)
interact with an environment involving other degrees of freedom whose properties
are not measured. \\

Decoherence is well-studied in the context of inflation \cite{Decoherence,Burgess:2014eoa,Nelson:2016kjm}.
Nelson \cite{Nelson:2016kjm} argued that the gravitational nonlinearities (from coupling between long-wavelength fluctuations and an environmental sector in the interaction action) provide a minimal mechanism for generating classical stochastic
perturbations from inflation via the decoherence. The best-suited framework to discuss about decoherence
of cosmological perturbations and the quantum to classical transition  is the Schr\"{o}dinger field theory. This picture is the natural framework to study the entanglement between the fields \cite{Martin-Martinez:2014gra} and can be used to study the entanglement effect on curvature power spectrum \cite{Albrecht:2014aga} and bispectrum using the interaction picture \cite{Rostami:2016ulp}.  There is no reason that one could not consider a more general initial
state such as an entangled one especially that inflation be an (low energy) effective theory of a fundamental theory such as quantum gravity or has multiple fields.\\

The purpose of the present article is to study the primary entanglement effect on the decoherence of fields  reduced density matrix which is in interaction with another field or independent mode functions. It will be shown that the primary entanglement has a significant role in decoherence of the system quantum state. We discuss that the existence of entanglement could couple  dynamical equations coming from Schr\"{o}dinger equation, and if someone wants to see no effect of entanglement parameter in decoherence then interaction terms in Hamiltonian can not be independent from each other. Generally, if we include  the primary entanglement  the independence of the interaction terms will be destroyed. \\

 The article is organized as follows. Section II review the decoherence in the quantum field theory. In the section III we present entanglement of fields in the inflationary background and study the possible interactions in the third order.  Section IV is devoted to  study the Schr\"{o}dinger equation for the entangled state and how the decoherence happen in this picture. Finally, we conclude with a discussion in section V.\\

\section{Decoherence in quantum physics}

One of the most important problems in quantum mechanics is the classicalization problem. The classicalization  means a process which transforms a quantum system to a classical one. The straightforward (and of course the most difficult) idea to unlock this problem is to replace the wave collapse assumption with a deterministic dynamical process which can describe how the collapse  happens. This idea entails a departure during measurement  from Born rule for instance collapse models \cite{Collapse}. One could try to find a non-fundamental solution for the case of statistical quantum systems. This method is not fundamental because implicitly includes Born rule. In despite it can describe how we find a statistical quantum system in a statistical classical system. In other words, it means how quantum probabilities change to classical probabilities.\\

What happens in relation between system and its environment  has been emerged in recent years a dramatic picture (which people like to call it measurement). This has been widely due to the attention of the phenomenon of decoherence. In this section, we review the decoherence concept.

 It is clear that the first requirement in the effect of environment on the system under study is an evolution of the state vector in the Schr\"{o}dinger picture, which creates a correlation between the system (like inflaton at the early universe) and states of environment (like other fields which affected on inflaton during inflation). Suppose that the system can be in various states labeled with an index $s$, while the environment can be in states labeled with an index $e$, such that the states of the total system in Hilbert space can be written in terms of a complete orthonormal basis of state vectors presented as $\Psi_{se}$. We assume that at $t = 0$ environment is placed in a initial state denoted $e = 0$, with the system in a general superposition of its states (in a subspace of the total Hilbert space ), so that the combined system would have an initial wave function as following 
 \begin{equation}
 \Psi (0)=\sum_{s} c_s \Psi_{s0} .
 \end{equation}
 
When we turn on an interaction between the system under study and the its correspond environment,  the combined system evolves in a time $t$ to $U \Psi (0)$,  where $U$ is the time evolution unitary operator $U = e^{- i t H  }$. To have an ideal decoherence, we need to choose the Hamiltonian $H$ to be in a way that the basis states  $\Psi_{s0}$  should evolve into states $U \Psi_{s0}  = \Psi_{s e_s}$ , with the index $s$ unchanged, and with $e_s$ labeling some definite state of the environment in a one to one correspondence with the state of the system under study, such that $e_s\neq e_{s'}$ if $s\neq s'$. For this, we just need 
\begin{equation}\label{03}
U_{s' e', s 0} = \delta_{s s'} \delta_{e' e_{s}}.
\end{equation}

It is possible always to choose the other elements of $U_{s' e', s e}$ with $e \neq 0$, to make the whole transformation unitary. For instance, in the case of $e\neq 0$, we can take this transformation as following 
\begin{equation}\label{04}
U_{s' e', s e} = \left \{ \begin{array}{rcl}
					\delta_{s s'} \mathcal{U}^{(s)}_{e e'}, & \mbox{for} & e' \neq e_{s'}	\\
					0, & \mbox{for} & e' = e_{s'}
\end{array}\right.
\end{equation}

where the matrix $\mathcal{U}^{(s)}$ has been constrained by the condition that, for all $e\neq 0$ and $e '\neq 0$,
\begin{equation}
\delta_{e e'} = \sum_{e''\neq e_{s}}{\mathcal{U}^{(s)}}^{*}_{e'' e'} \mathcal{U}^{(s)}_{e'' e}.
\end{equation}
These conditions thus simply require that  $\mathcal{U}^{(s)}$  are unitary matrices, and since they are not subject to any  other constraints, one can establish any number of matrices which satisfy this condition.

After the system under study and the environment have interacted, the total system would be found in following superposition 
\begin{equation}\label{01}
U\Psi(0) = \sum_s c_s \Psi_{s e_s},
\end{equation}
which is an entangled state of system and environment created by interaction. We have not yet a decoherence, because the combined system is still in a pure state and we just see a definite superposition of the basis. Based on Born rule the system must make a transition during the measurement to one or other of these states, with probabilities ${|c_s|}^2$. Here by classical state we mean  the favored states produced by measurement (interaction between system and environment) which the system under study goes to them. Zurek identified such sates with the name of “pointer states.” 

After this introduction we are ready to ask why we see most of systems around us classical? 
The answer has to do with the phenomenon of decoherence. This happens because any specific environment  will always be subject to tiny noises which could rises  the environment number of degrees of freedom extremely. These perturbations could not by themselves change one classical state into another one. We can investigate this issue in two  equivalent looks. The  decoherence converts Eq. (\ref{01}) as following
\begin{equation}
\sum_s c_s \Psi_{s e_s} \longrightarrow \sum_s  \exp(i \phi_s)\  c_s \Psi_{s e_s}, 
\end{equation}
where the $\phi_s$ are randomly fluctuating phases. Consequently, when we take account the expectation values the interferences between different terms in the above superposition would average to zero, and the expectation value of any observer operator $ A$ gives 
\begin{equation}\label{02}
\overline{\langle A \rangle } = \sum_s {|c_s|}^2 \left( \Psi_{s e_s} , A  \Psi_{s e_s} \right),
\end{equation}
with the bar over the expectation value indicating that it is averaged over the
phases $\phi_s$. Here we see that the expectation value of $A$ is just given by a classical distribution. One may note that this is not really a solution for measurement problem because we have used Born rule in (\ref{02}). \\

One can also indicate this  phenomenon in another way (and equivalent to the former). To see it better, we go to the usual ket-bra notation. Suppose that $|E\rangle$ and $|S_i\rangle$ be states of environment and system respectively. Here we have assumed that $|S_i\rangle$ states establish a orthonormal subspace. It is clear that the interaction defined in Eq. (\ref{03}) and (\ref{04}) takes the combined system at $t_0$ to any later time as following
\begin{equation}
\vert E(t_0)\rangle \vert S_i (t_0) \rangle \longrightarrow \vert E_i (t)\rangle \vert S_i (t) \rangle.
\end{equation}
Then if the environment acts as during an ideal measurement we will have 
\begin{equation}\label{05}
\langle E_i (t) \vert E_j (t) \rangle \approx \delta_{ij}.
\end{equation}
Note that this would  happen when there are many number of degrees of freedom for the environment. Now an initially coherent superposition of system goes to an entanglement state when time pasts as (\ref{01})
\begin{equation}\label{06}
\vert E(t_0)\rangle \left( \sum_i c_i \vert S_i (t_0) \rangle\right)  \longrightarrow \sum_i c_i \vert E_i (t)\rangle \vert S_i (t) \rangle.
\end{equation}
Therefore, to see how decoherence comes across it is enough to find reduced density matrix of system under study. If this matrix be  respect $\vert S_i\rangle$ then we have a classical distribution. Using of Eq. (\ref{05}),  the components of reduced density matrix become 
\begin{equation}
\rho_R (S_i, S_j) \approx {|c_i|}^2 \delta_{ij}.
\end{equation}  
Then the effect of such kind of interactions is to  eliminate off diagonal components of a density matrix.

It would be useful to translate above discussion to a scalar field theory in Schr\"{o}dinger picture.
Against Heisenberg picture for a field theory in  which one works with a specific Fock space, in Schr\"{o}dinger picture we use wave functional to describe what are happening in a quantum system. To do that, suppose we have a scalar theory for $\phi(x)$. Then in Schr\"{o}dinger picture, we have to find a basis for this theory. Thus, this is natural to choose the eigenstates of the operator $\hat{\phi}(x)$ as a suitable basis which has been defined as following
\begin{equation}
\hat{\phi}(x) \vert \phi(x)\rangle = \phi(x) \vert \phi(x)\rangle.
\end{equation}
 Now an arbitrary state could be represented as a superposition of the field eigenstates, 
 \begin{equation}\label{07}
 \vert \Psi_{\phi}\rangle = \sum_{\phi(x)} \Psi[\phi(x)]\  \vert \phi(x)\rangle,
 \end{equation}
 which $\Psi[\phi(x)]$ has the role of wave functional. Note that the above summation is functional integration and we write down it formally. To establish a decohering system it's enough to treat like what we did in Eq. (\ref{06}),
 \begin{equation}
 \vert \Psi_{E}\rangle \vert \phi(x)\rangle \longrightarrow \left( \vert \Psi_{E} |_{\phi(x)}\rangle \right) \vert \phi(x)\rangle.
 \end{equation}
 Therefore, if one establish a simple combined system at the initial state $\vert \Psi_{E}\rangle  \vert \Psi_{\phi}\rangle$, the one could easily show that the correspond reduced density matrix of $\phi(x)$ becomes
 \begin{equation}\label{e8}
 \rho_R [\phi(x), \phi^{'} (x)] = \Psi_{\phi}[\phi(x)] \Psi_{\phi}^{*}[\phi^{'}(x)] \sum_{E} \left( \Psi_{E}[E]\vert_{\phi(x)}\right) \left( \Psi^{*}_{E}[E]\vert_{\phi^{'} (x)} \right).
 \end{equation}
 Now if the interaction leads the summation term in the above equality to zero, then we will have decoherence. To this end, we are tracking some interaction terms such that satisfy this condition.\\

\section{Schorodinge equation for entangled fields} 

In this section we  expand the Schr\"{o}dinger field theory during inflation for a combined system including scalar field $\phi(x)$ or a special degree of freedom (as the main system) and another field like $\chi(x)$ or the rest of degrees of freedom (as the environment). What we are looking for is wave function of the combined system. Once we find it, then all information of environment and system under study would be obtained and then we would be able to see if the system can experience decoherence or not. Here the field $\phi(x)$ has the role of fluctuation of inflaton and the field $\chi(x)$ is another field which could exist during inflation but has no role in dynamics of inflation. Nevertheless, depending on initial state, the entanglement between this field and inflaton could appear in power spectrum and bispectrum of the inflationary universe \cite{Bolis:2016vas}.  We are interested in to know, what's happened for the wave function when one start from an entangled state of system and environment. We would like to emphasis that the following is not just for two fields or inflation theory or different length modes but could even use for scalar and tensor modes interactions. 

The total wave function would evolve according to the Schr\"{o}dinger equation

 \begin{equation}\label{07}
 i \frac{d}{dt} \Psi[E, S] = H[E, S; t] \Psi[E,S]
 \end{equation}
with the time-depended Hamiltonian $H(t)$.  We assume that the interaction between system and environment could be treated perturbatively. In our case, this is totally reasonable because we are working with fluctuations of fields. The accuracy of such calculation would be valid up to the third order of fluctuations. This Hamiltonian includes free terms of fields and the interaction part
 \begin{equation}\label{08}
 H[E,S] = H_{0}[E,S] + H_{int}[E, S],
 \end{equation}
 which $H_{int}$ is the interaction between system and environment \footnote{For instance in the case which the long wavelength modes and short ones have the role of system and environment respectively, this interaction comes from the cubic terms in the perturbed action \cite{Maldacena:2002vr,Nelson:2016kjm}. Moreover, these interaction terms can come from the extension of the standard model of particle physics or moduli of compactification in string theory.}. Here the free Hamiltonian $H_0$ includes a kinetic term with Fourier transformation
 \begin{equation}\label{09}
 H_{k}[S]= \frac{1}{2} \int \frac{d^{3} p}{(2\pi)^3} f_{s}(\tau) \pi_{p}[S] \pi_{p}^{*}[S]
 \end{equation}
which the conjugate momentum is defined by
 \begin{equation}
 \pi_{p}[S]=-i \frac{\delta}{\delta \phi_{p}}.
 \end{equation}
Here $f_s$ depends on the geometry of space time and the kind of fields. For example in case of de Sitter space time for a free scalar field action one find it $f_s = \frac{1}{2 a(\tau)}$ and it is independent of fields although in general hyper-globally space time it couldn't be true \cite{Townsend:1997ku}.  To solve Eq. \eqref{07}  perturbatively, one needs to know the solution of the free part in Eq. (\ref{08}). The general solution $\Psi_{en}[E, S]$ for  two independent free scalar fields, should satisfy
 \begin{equation}\label{10}
 i \frac{d}{dt} \Psi_{en}[E, S] = H_{0}[E, S; t] \Psi_{en}[E,S].
 \end{equation}
  One could do it just by an anzats for ground state
  \begin{equation}\label{11}
 \Psi_{en}= N_{en}(\tau) \exp \left[ -\int \frac{d^{3} k}{(2\pi)^3}\left( A_{k}(\tau) \phi_{k}\phi_{-k} + B_{k}(\tau) \chi_{k} \chi_{-k} + 2 C_{k}(\tau) \phi_{k} \chi_{-k}\right) \right],
 \end{equation}
 and finds some definite differential equations for $A$, $B$, and entangled parameter $C$ and finally solve them with a initial condition, suitable for a inflation theory. One notes that, this form of wave function\footnote{The entanglement is put in here when $C(\tau_{0}) \neq 0$ at the beginning. In fact, using Schr\"{o}dinger equation for free part of combined Hamiltonian, one could  find the simple equation $\frac{C_{k}^{'}}{C_{k}}=\frac{(A_k + B_k)}{a^{2} (\tau)}$. The non-vanishing value of entangled parameter $C_{k}$ in the start of inflation, gives non-zero value of it when inflation goes on \cite{Bolis:2016vas}. It maybe useful to know that, the existence of $C_{k}$ in Schr\"{o}dinger picture is equivalence to  making an entanglement with Bogoliubov transformation of vacuum in Heisenberg picture \cite{Kanno:2015ewa}.  } is invariant under rotations and spatial translations. This solution would be Gaussian because of quadratic form of the free Hamiltonian. Now, to solve Hamiltonian including interactions of cubic terms, we suppose there is solution as following
 \begin{equation}\label{12}
i \Psi_{en} \dot{\Psi}_{ng} = \left( H_{k}[S] + H_{k}[E] + H_{int}[E, S]\right) \Psi_{en} \Psi_{ng}.
 \end{equation}
Here we see that the effect of interactions appear as a non-Gaussian part $\Psi_{ng}$ in the wave function at the ground state. Note that we have just used kinetic parts of Hamiltonian.  When one uses Eq. (\ref{10}), the potential terms of free Hamiltonian would be canceled from RHS of Eq. (\ref{12}). Using of Eq. (\ref{09}), one can find the effective term of kinetic parts in Eq. (\ref{12}) 
 \begin{equation}\label{13}
 H_{k}[S]  \Psi_{en} \Psi_{ng} \rightarrow \frac{1}{2} \int \frac{d^{3}p}{(2\pi)^3} f_{s}(\tau)\{ \frac{\delta \Psi_{en}}{\delta \phi_{-p}} \frac{\delta \Psi_{ng}}{\delta \phi_{p}} \}
 \end{equation}
and the same for environment. One can again find a good anzats for the non-Gaussian part of wave function as
 \begin{equation}\label{14}
 \Psi_{ng}=\exp \int_{k^{'}, k, p} \left( \phi_p \chi_{k} \chi_{k^{'}} F_{k k^{'} p} +  \phi_p \phi_{k} \phi_{k^{'}} M_{k k^{'} p} +  \chi_p \phi_{k} \phi_{k^{'}} N_{k k^{'} p} +  \chi_p \chi_{k} \chi_{k^{'}} Q_{k k^{'} p} \right)
 \end{equation}
which the dynamical coefficients $N, M, F$ and $Q$ must be find by Schr\"{o}dinger equation \footnote{One may want to know why the normalization factor has not been considered in (\ref{14}). The reason is that the dynamical equation for normalization is related to zero order of perturbation and so is irrelevant here. In other words, this consideration has not any effect in derivation of (\ref{16}- \ref{19})}. In the above integral we use the usual convention $\int_{k, k^{'}, p} \equiv \int \frac{dk^3}{(2\pi)^{3}} \frac{{dk'}^3}{(2\pi)^{3}} \frac{dp^3}{(2\pi)^{3}} \delta (k+k^{'}+p)$. Also it is convenient to write down action of the cubic interaction on the solution in Fourier space like
\begin{equation}\label{15}
H_{int}  \Psi_{en} \Psi_{ng} =  \int_{k^{'}, k, p}\left[ \mathcal{H}^{(1)}_{k k^{'} p} \phi_{k} \phi_{k^{'}} \phi_{p} + \mathcal{H}^{(2)}_{k k^{'} p} \chi_{k} \chi_{k^{'}} \phi_{p} + \mathcal{H}^{(3)}_{k k^{'} p} \phi_{k} \phi_{k^{'}} \chi_{p} + \mathcal{H}^{(4)}_{k k^{'} p} \chi_{k} \chi_{k^{'}} \chi_{p}\right]  \Psi_{en} \Psi_{ng},
\end{equation}
which the integration is just eigenvalue of $H_{int}$ operator. Substituting Eq. (\ref{14}), (\ref{15}) and (\ref{11}) in (\ref{12}) and using this fact that all cubic multiplication of fields in Schr\"{o}dinger equation are independent, one would get four coupled first order differential equations for unknown dynamical variables in non-Gaussian part of wave function \footnote{Here we have used conformal time instead of cosmological time by substituting $\frac{d}{dt}= -H \frac{d}{d\tau}$} Eq. (\ref{14})  
\begin{eqnarray}\label{16}
 & -i H \tau \dot{F}_{k k^{'} p} = \nonumber \\  & f_{s} (p,\tau) A_{p}(\tau) F_{k k^{'} p} + 2 f_{s} (k,\tau) C_{k}(\tau) N_{k p k^{'} } + 2 f_{e} (k^{'},\tau) B_{k^{'}}(\tau) F_{k k^{'} p} + 3 f_{e} (p,\tau) C_{p}(\tau) Q_{k k^{'} p} + \mathcal{H}^{(2)}_{k k^{'} p}
\end{eqnarray}
which comes from $\chi^2 \phi$ coefficients in Schr\"{o}dinger equation, and
\begin{equation}\label{17}
-i H \tau \dot{M}_{k k^{'} p} = 3 f_{s} (p,\tau) A_{p}(\tau) M_{k k^{'} p} + f_{e} (p,\tau) C_{p}(\tau) N_{k k^{'} p} + \mathcal{H}^{(1)}_{k k^{'} p}
\end{equation}
is the coefficient of $\phi^3$, and
\begin{equation}\label{18}
-i H \tau \dot{Q}_{k k^{'} p} =  f_{s} (p,\tau) C_{p}(\tau) F_{k k^{'} p} + 3 f_{e} (p,\tau) B_{p}(\tau) Q_{k k^{'} p} + \mathcal{H}^{(4)}_{k k^{'} p}
\end{equation}
is related to the term of $\chi^3$, and finally
\begin{eqnarray}\label{19}
& -i H \tau \dot{N}_{k k^{'} p} =  \nonumber \\ & 3 f_{s} (p,\tau) C_{p}(\tau) M_{k k^{'} p} + 2 f_{s} (k^{'},\tau) A_{k^{'}}(\tau) N_{k k^{'} p } + 2 f_{e} (k,\tau) C_{k}(\tau) F_{k p k^{'}} +  f_{e} (p,\tau) B_{p}(\tau) N_{k k^{'} p} + \mathcal{H}^{(3)}_{k k^{'} p},
\end{eqnarray}
which is the coefficient of $\phi^2 \chi$. In Eq. (\ref{16}-\ref{19}) we've used dote as conformal time derivation. Solving these equations we able to talk about all quantum effect on inflaton perturbations like decoherence. In the next section, we will show that just Eq.  (\ref{16}) is related to decoherence and try to find some exact solutions for these equations. In general one has to solve these kind of equations numerically which we left it for next papers.\\

We end this section with understanding this question that, why  the entanglement parameter is related to interaction parts. To answer this question one may look at the path integral method. We know that one could relate the propagator to wave function from path integral method to Schr\"{o}dinger  picture version of quantum mechanics
\begin{equation}
\Psi \leftrightarrow \int \mathcal{D} X \ \exp\left(i S\right).
\end{equation}  
Now, if we insert the interaction part to the exponent in RHS, then the LHS should be modified and vice versa. Because the RHS would be changed exponentially, it is reasonable, the LHS be modified as the same.\\

\section{Decoherence from entanglement} 
In this section we want to find some exact solution of Eq. (\ref{16}-\ref{19}) and investigate the phenomenon of decoherence related to the form of interactions. As mentioned before, it seems difficult to solve this coupled system of differential equations although one could try for numerical ones. Here we shall check two especial exact solutions of them.\\

\subsection{Non-entangled Case}

One of the interesting case is non-entangled states. In this case since $C=0$ for all modes, system and environment are not correlated to each other at early time. Now, we analyze the above coupled system of differential equations and understand more about decoherence in such theories.
When we justify entangled parameter to zero, this coupled system of equations is transformed to decoupled one and so the solution would be pretty easy. At first, we would like to focus on Eq.  (\ref{16}) 
\begin{equation}\label{20}
i H \tau \dot{F}_{k k^{'} p} + g(\tau; k, k^{'}, p) H\tau F_{k k^{'} p} + \mathcal{H}^{(2)}_{k k^{'} p}=0
\end{equation}
where $g\equiv f_s A + 2 f_e B$. To solve it, we need an initial condition. Actually we are interested in theories which interactions are active during inflation and have no effect at early time. With this assumption, it would be reasonable to assume $F(\tau_0)=0$. Therefore, the solution is
\begin{equation}\label{21}
F_{k k^{'} p}(\tau) = i \int_{\tau_0}^{\tau} \frac{d\tau^{'}}{H\tau^{'}}  \mathcal{H}^{(2)}_{k k^{'} p}(\tau^{'}) \exp \left[ i \int_{\tau^{'}}^{\tau} d\tau^{''} g(\tau^{''}; k, k^{'}, p)\right].
\end{equation}
This relation would be more simple in some cases. For example, in de Sitter space suppose the theory  in which $\mathcal{H}^{(2)}$ is proportional to $a^n$ with $n > 0$ for scale factor \cite{Nelson:2016kjm}. If the coupling between system and environment is weak enough such that the density matrix remains close to Gaussian, then the real part of $F$ doesn't grow at late time. In other words, when $\tau^{'}$ is close to $\tau$, the exponent part in Eq. (\ref{22}) is negligible. So for late time we have just imaginary part of this as following
\begin{equation}\label{22}
\lim_{\tau \rightarrow 0}Im F_{k k^{'} p}(\tau) = \int_{\tau_0}^{\tau} \frac{d\tau^{'}}{H\tau^{'}}  \mathcal{H}^{(2)}_{k k^{'} p}(\tau^{'}).
\end{equation}
Now, let's come back to decoherence phenomenon. The summation in the reduced density matrix Eq.  (\ref{e8}) is proportional to 
\begin{eqnarray}
& \sum_{E} \left( \Psi_{E}[E]\vert_{\phi(x)}\right) \left( \Psi^{*}_{E}[E]\vert_{\phi^{'} (x)} \right) \propto \int \mathcal{D}\chi \vert \Psi_{en}[E, S]\vert^{2} \exp \left[ \int_{k, k^{'}, p} \chi_{k} \chi_{k^{'}}\left( \phi_{p} F_{k k^{'} p} + \phi^{'}_{p} F^{*}_{k k^{'} p}\right) \right] = \nonumber \\ &
 \langle \exp \left[ i \int_{k, k^{'}, p} \chi_{k} \chi_{k^{'}} \Delta\phi_{p} Im(F_{k k^{'} p}) \right] \rangle
\end{eqnarray}
where $\Delta\phi_{p}=\phi_{p} - \phi'_{p}$, and this equation appears as a average value of the exponential on environment degrees of freedom. Thus, based on Riemann's integration theorem, if imaginary part of $F$ be large then the off-diagonal elements of density matrix go to zero and decoherence would happen. From here, we can see $F$ is the most important term which decoherence phenomenon is related to. This happens because $\Delta \phi$ has been coupled only with $F$.

Although the dynamics of $Q$, $N$ and $M$ have no effect in decoherence but for a complete description, we shall solve them here. Because all these differential equations are decoupled and the same, the solutions are like the solution of $F$ if we choose the same initial conditions:
 \begin{equation}\label{23}
M_{k k^{'} p}(\tau) = i \int_{\tau_0}^{\tau} \frac{d\tau^{'}}{H\tau^{'}}  \mathcal{H}^{(1)}_{k k^{'} p}(\tau^{'}) \exp \left[ i \int_{\tau^{'}}^{\tau} d\tau^{''} m(\tau^{''}; k, k^{'}, p)\right],
\end{equation}
where $m\equiv 3 f_s A$. For $N$ we have 
\begin{equation}\label{24}
N_{k k^{'} p}(\tau) = i \int_{\tau_0}^{\tau} \frac{d\tau^{'}}{H\tau^{'}}  \mathcal{H}^{(3)}_{k k^{'} p}(\tau^{'}) \exp \left[ i \int_{\tau^{'}}^{\tau} d\tau^{''} n(\tau^{''}; k, k^{'}, p)\right],
\end{equation}
which $n\equiv 2 f_s A + f_e B$, and finally
\begin{equation}\label{25}
Q_{k k^{'} p}(\tau) = i \int_{\tau_0}^{\tau} \frac{d\tau^{'}}{H\tau^{'}}  \mathcal{H}^{(4)}_{k k^{'} p}(\tau^{'}) \exp \left[ i \int_{\tau^{'}}^{\tau} d\tau^{''} q(\tau^{''}; k, k^{'}, p)\right],
\end{equation}
where $q\equiv 3f_e B $. The significance of these terms is in calculation related to bi-spectrum but not used in the decoherence of density matrix or spectrum of CMB. In fact one could make an standard theory for inflation just by choosing $\mathcal{H}^{(1)}= \mathcal{H}^{(3)}= \mathcal{H}^{(4)}=0$. In such theory, $N$, $Q$ and $M$ are equal to zero and $F$ has a nonzero value which can give a contribution to bispectrum of CMB. 
\subsection{Entangled Case}
In this part of paper, we investigate a theory in which decoherence happens as  before case with this difference that the theory has the entangled initial state. In the previous case we saw that if there is no entanglement in combined system then there wouldn't be any correlation between other interaction terms. Here we will find a solution in the presence of entanglement such that interaction terms are related to entanglement variable $C$. 

To find such solution we choose $N=Q=0$. Therefore, Eq. (\ref{16}) which is responsible of decoherence phenomenon would be the same as before and decoherence will be happened like the case $C=0$. One can easily see that the solution for $M$ is the same as Eq. ($\ref{23}$) but there are two consistency relations for  $\mathcal{H}^{(3)}$ and $\mathcal{H}^{(4)}$ as following 
\begin{eqnarray}\label{26}
3i f_{s} (p,\tau) C_{p}(\tau)   \int_{\tau_0}^{\tau} \frac{d\tau^{'}}{H\tau^{'}}  \mathcal{H}^{(1)}_{k k^{'} p}(\tau^{'}) \exp \left[ i \int_{\tau^{'}}^{\tau} d\tau^{''} m(\tau^{''}; k, k^{'}, p)\right] + \nonumber \\  2i f_{e} (k,\tau) C_{k}(\tau)   \int_{\tau_0}^{\tau} \frac{d\tau^{'}}{H\tau^{'}}  \mathcal{H}^{(2)}_{k k^{'} p}(\tau^{'}) \exp \left[ i \int_{\tau^{'}}^{\tau} d\tau^{''} g(\tau^{''}; k, k^{'}, p)\right] = - \mathcal{H}^{(3)}_{k k^{'} p}
\end{eqnarray}
and
\begin{equation}\label{27}
i f_{s} (p,\tau) C_{p}(\tau) \int_{\tau_0}^{\tau} \frac{d\tau^{'}}{H\tau^{'}}  \mathcal{H}^{(2)}_{k k^{'} p}(\tau^{'}) \exp \left[ i \int_{\tau^{'}}^{\tau} d\tau^{''} g(\tau^{''}; k, k^{'}, p)\right] =- \mathcal{H}^{(4)}_{k k^{'} p}.
\end{equation}
These equations imply that two of four interactions are not independent and for example once $\mathcal{H}^{(1)}$ and $\mathcal{H}^{(2)}$ are given, the others would be completely defined.
One may know if Eq. (\ref{16}-\ref{19}) have solutions such that affect on decoherence process. In other words, the question arise as to whether there is a solution in which decoherence depends on entanglement. Generally the answer is positive and this can be seen by choosing $N=0$ and $Q\neq 0$. Here, one should note that the entanglement parameter $C$ is independent from the magnitude of slow-roll parameters in case of inflationary universe example. This happens because $C$  just depends  on ratio of interactions.   Now, Eq.  (\ref{16}) is no longer independent of other equations and should  be solved again with entanglement variable $C$. Here we can consider $\mathcal{H}^{(4)}$ as a independent term and use equation (\ref{19}) to find $F$ as 
\begin{eqnarray}\label{28}
 & F_{k p k^{'}} = \nonumber \\  & -\frac{i}{2 f_{e} (k,\tau) C_{k}(\tau)}3 f_{s} (p,\tau) C_{p}(\tau)  \int_{\tau_0}^{\tau} \frac{d\tau^{'}}{H\tau^{'}}  \mathcal{H}^{(1)}_{k k^{'} p}(\tau^{'}) \exp \left[ i \int_{\tau^{'}}^{\tau} d\tau^{''} m(\tau^{''}; k, k^{'}, p)\right]  -\frac{1}{2 f_{e} (k,\tau) C_{k}(\tau)} \mathcal{H}^{(3)}_{k k^{'} p}.
\end{eqnarray} 
We see here in this case decoherence is completely affected by $\lambda_{k,p}\equiv \frac{C_{p}(\tau)}{C_{k}(\tau)}$. In fact if $\lambda$ is very small then decoherence happens weakly but if this is very large then decoherence happens rapidly based on how  the behavior of $\mathcal{H}^{(1)}$ is.
There are two important key points: first, the last term in the right hand side Eq. (\ref{28}) has no effect on decoherence because the imaginary part of $F$ plays role in the decoherence; second, in contrast to the Eq. (\ref{22}) which the imaginary part of $F$ is proportional to the integration of  $ \mathcal{H}^{(2)} $, the imaginary part of $F$ is related to the integration of $ \mathcal{H}^{(1)} $. As a result, $ \mathcal{H}^{(1)} $ has to satisfy the same condition needed for $ \mathcal{H}^{(2)} $ to have decoherence \cite{Burgess:2014eoa,Nelson:2016kjm}.\\

\section{Conclusion}

Our goal of this work was to investigate the decoherence of a field (inflation in cosmology) which is not alone in the universe and it can be in interaction with environment or any other independent modes function. There is no reason one could not consider a more general initial
state with including the entanglement between field and modes. Especially that inflation be an (low energy) effective theory of a fundamental theory such as quantum gravity or has multiple fields.

We have two types of entanglements: first, primary entanglement which comes from initial state of combined system including environment and the system under study; second, late entanglement come cross during interaction system and environment. To have decoherence during the interaction the late entanglement is necessary.\\ 

Let us conclude with some remarks:
\begin{itemize}
	\item In this paper we showed that the primary entanglement has a significant role in decoherence of the system quantum state.
	
	\item It was shown that if there is no primary entanglement in combined system then the interaction terms which can be responsible for late entanglement are independent. In this case, the interaction term, $H_{int} \sim   \int_{k^{'}, k, p} \mathcal{H}^{(2)}_{k k^{'} p} \chi_{k} \chi_{k^{'}} \phi_{p} $,   contribute to the decoherence.

	\item If we have primary entanglement $C \neq 0$ there is a solution of the Schr\"{o}dinger wave equation, $N=Q=0$, the interaction terms are not independent. If we have two of the interaction terms then we can find the others. It is commonly believed that we can add any interaction term between combined system, but this result show that the primary entanglement limit us to choose the interaction term. In this case the decoherence happens like the case of no primary entanglement, $C = 0$.

	\item We also find another exact solution $C,Q \neq 0$, $N=0$, in which the decoherence process is directly affected by entanglement parameter $C$. There are two important key points here: first, the last term in the right hand side Eq. (\ref{28}) has no effect on decoherence because the imaginary part of $F$ plays role in the decoherence; second, in contrast to the Eq. (\ref{22}) which the imaginary part of $F$ is proportional to the integration of  $H_{int} \sim   \int_{k^{'}, k, p} \mathcal{H}^{(2)}_{k k^{'} p} \chi_{k} \chi_{k^{'}} \phi_{p} $, the imaginary part of $F$ is related to the integration of $H_{int} \sim   \int_{k^{'}, k, p} \mathcal{H}^{(1)}_{k k^{'} p} \phi_{k} \phi_{k^{'}} \phi_{p} $. As a result, $ \mathcal{H}^{(1)} $ has to satisfy the same condition needed for $ \mathcal{H}^{(2)} $ to have decoherence \cite{Burgess:2014eoa,Nelson:2016kjm}.

		\item The dependency between the primary entanglement and the interaction terms can have can have a teleological interpretation. Suppose that the semi-classical picture of inflation theory is an effective low energy theory of a universal quantum gravity theory (UQGT). Therefore, both primary entangled states and the interaction terms  emerge from the low energy limitation from UQGT. From this prospective, the entanglement and the interaction terms can not be independent.
		
	\item And at the end we should to emphasis that the difference between theories with different interactions could  appear not only their two-points correlation functions at early time but also three point functions are needed. The contributions of these three-points functions come from non-Gaussian part of theories which now is related to entanglement parameter $C_{k}$ in the general solution.
	
\end{itemize}

Several directions for future research exist; One can use a multi fields model (for example two scalar fields  or tensor-scalar field models) like with primary entanglement, to look at the decoherence rate of wave function in the super horizon and verify whether this entanglement delay the classicalization or not. Even for single field inflation, one can look at third order action with different coupling between independent modes and look at wave density matrix  decoherence in presence of primary entanglement.  It would be also interesting to study the dynamics of the entangled  state in phase space. With calculating the wave function's Wigner function, we can understand the  coherence lengths and squeezing at late times, and whether the diagonal matrix elements evolve according to the standard Fokker-Planck equation of Starobinsky’s stochastic inflation. One can also study  the entanglement effects on the redundant records of long wavelength perturbation during inflation to study the squeezing of the quantum stats \cite{Nelson:2017pmc}. \\

{\bf Acknowledgments:}
\\

We would like to thank M.M. Sheikh-Jabbari and Hassan Firouzjahi for useful comments.
\\

\end{document}